\documentclass[10pt,conference]{IEEEtran}

\usepackage[dvips]{color}
\usepackage{tabu}
\usepackage{comment}
\usepackage{todonotes}
\usepackage{epsf}
\usepackage{epsfig}
\usepackage{times}
\usepackage{epsfig}
\usepackage{graphicx}
\usepackage{bbold}
\usepackage{mathtools}
\usepackage{mathrsfs}
\usepackage{amssymb,amsmath}
\usepackage{pdfpages}
\usepackage{epstopdf}
\usepackage{dsfont}
\usepackage{lettrine} 
\usepackage{amsmath,epsfig,amssymb,algorithm,algpseudocode,amsthm,cite,url}

\usepackage[justification=centering]{caption}
\usepackage{subcaption}
\usepackage{bbm}
\usepackage{algorithm}
\usepackage{algpseudocode}
\usepackage{algpseudocode}
\algnewcommand{\LineComment}[1]{\State \(\triangleright\) #1}
\usepackage{csquotes}
\usepackage{amsmath,amssymb}

\DeclareFontFamily{OMX}{yhex}{}
\DeclareFontShape{OMX}{yhex}{m}{n}{<->yhcmex10}{}
\DeclareSymbolFont{yhlargesymbols}{OMX}{yhex}{m}{n}
\DeclareMathAccent{\wideparen}{\mathord}{yhlargesymbols}{"F3}
\usepackage{verbatim}
\usepackage{amsmath,amssymb}

\usepackage{verbatim}

\newtheorem{theorem}{\bf Theorem}
\let\oldtheorem\theorem
\renewcommand{\theorem}{\oldtheorem\normalfont}

\newtheorem{proposition}{\bf Proposition}
\let\oldproposition\proposition
\renewcommand{\proposition}{\oldproposition\normalfont}
\newtheorem{lemma}{\bf Lemma}
\let\oldlemma\lemma
\renewcommand{\lemma}{\oldlemma\normalfont}
\newtheorem{example}{Example}
\let\oldexample\example
\renewcommand{\example}{\oldexample\normalfont}

\newtheorem{definition}{\bf Definition}
\let\olddefinition\definition
\renewcommand{\definition}{\olddefinition\normalfont}
\newtheorem{remark}{Remark}
\let\oldremark\remark
\renewcommand{\remark}{\oldremark\normalfont}

\usepackage{setspace}	
\allowdisplaybreaks

\IEEEoverridecommandlockouts
\begin{document}
\title{\huge Performance Analysis of Integrated Sub-6~GHz-Millimeter~Wave
	 Wireless Local Area Networks}
\author{
\authorblockN{Omid Semiari$^{1}$, Walid Saad$^{2}$, Mehdi Bennis$^3$, and Merouane Debbah$^4$}\\\vspace*{-1em}
\authorblockA{\small $^{1}$\textcolor{black}{Department of Electrical Engineering, Georgia Southern University, Statesboro, GA, USA, Email: \protect\url{osemiari@georgiasouthern.edu}}\\
	$^{2}$Wireless@VT, Bradley Department of Electrical and Computer Engineering, Virginia Tech, Blacksburg, VA, USA, Email: \protect\url{walids@vt.edu}\\
\small $^3$ Centre for Wireless Communications, University of Oulu, Finland, Email: \url{bennis@ee.oulu.fi}\\
\small $^4$ Mathematical and Algorithmic Sciences Lab, Huawei France R\&D, Paris, France, Email: \url{merouane.debbah@huawei.com}
}\vspace*{-2.3em}
   \thanks{This research was supported by the U.S. National Science Foundation under Grants CNS-1460316 and CNS-1526844, and by the ERC Starting Grant 305123 MORE.
   	}%
  }
%
\maketitle
\begin{abstract}
Millimeter wave (mmW) communications at the 60 GHz unlicensed band is seen as a promising approach for boosting the capacity of wireless local area networks (WLANs). If properly integrated into legacy IEEE 802.11 standards, mmW communications can offer substantial gains by offloading traffic from congested sub-6 GHz unlicensed bands to the 60 GHz mmW frequency band. In this paper, a novel medium access control (MAC) is proposed to dynamically manage the WLAN traffic over the unlicensed mmW and sub-6 GHz bands. The proposed protocol leverages the capability of advanced multi-band wireless stations (STAs) to perform fast session transfers (FST) to the mmW band, while considering the intermittent channel at the 60 GHz band and the level of congestion observed over the sub-6 GHz bands. The performance of the proposed scheme is analytically studied via a new Markov chain model and the probability of transmissions over the mmW and sub-6 GHz bands, as well as the aggregated saturation throughput are derived. In addition, analytical results are validated by simulation results. Simulation results show that the proposed integrated mmW-sub 6 GHz MAC protocol yields significant performance gains, in terms of maximizing the saturation throughput and minimizing the delay experienced by the STAs. The results also  shed light on the tradeoffs between the achievable gains and the overhead introduced by the FST procedure. 
 \vspace{-0cm}
\end{abstract}
\section{Introduction} \vspace{-0cm}
Advanced wireless stations (STAs) are capable of supporting multiple wireless local area network (WLAN) standards, including legacy IEEE 802.11 over the \emph{sub-6 GHz (microwave) unlicensed bands}, as well as IEEE 802.11ad over the \emph{$60$ GHz millimeter wave (mmW) band}\cite{7120076}. These modern STAs, also known as tri-band WiGig devices, can potentially benefit from high capacity mmW communications along with flexible, simple, and more reliable networking at the sub-6 GHz bands. Reaping the benefits of such a multi-band WLAN capability is contingent upon adopting new medium access control (MAC) protocols that can support flexible and dynamic traffic scheduling over the aggregated mmW--microwave ($\mu$W) unlicensed frequency bands
\footnote{Hereinafter, $\mu$W unlicensed band refers to either $2.4$ GHz, $5$ GHz, or both.}. Such promising integrated mmW-$\mu$W protocols also provide substantial motivation to revisit the existing MAC solutions for traditional, yet important challenges of WLANs. One such problem is the excessive delay at the contention-based medium access of the IEEE 802.11 standards that prevents WLANs to meet the stringent quality-of-service (QoS) requirements of emerging technologies, such as smart home applications\cite{1204546,1357025}.


The performance of IEEE 802.11 MAC protocols has been thoroughly studied in the literature \cite{840210,4100720,4107953,4288745,1325887}. The seminal work of Bianchi in \cite{840210} presents a comprehensive analysis for the performance of the distributed coordination function (DCF) of the IEEE 802.11. The authors in \cite{4100720} study the modeling and performance analysis of IEEE 802.11 DCF  in unsaturated scenarios with heterogeneous traffic arrival rates for STAs.  In \cite{4107953}, the authors propose a cooperative MAC protocol that leverages spatial diversity across the network to increase system throughput. The authors in \cite{4288745} study the performance of enhanced-DCF (EDCF) for IEEE 802.11e standard. Moreover, the work in \cite{1325887} and references therein propose different MAC protocols to improve QoS in IEEE 802.11. Although interesting, the body of work in \cite{840210,4100720,4107953,4288745,1325887} solely focuses on the WLAN standards at the $\mu$W unlicensed bands.

However, mmW communications over the 60 GHz unlicensed band is one of the key enablers to support emerging bandwidth-intensive technologies, such as virtual reality, in  WLANs\cite{11ad,7870651,6451302,5956194}. In fact, the large available bandwidth at 60 GHz mmW band allows STAs to potentially achieve higher data rates, compared with the data rates at the sub-6 GHz unlicensed $\mu$W bands. However, mmW links are inherently intermittent, due to extreme susceptibility of mmW signals to blockage \cite{7929424}. In addition, the challenges of bidirectional transmissions at the 60 GHz band, such as deafness, increase the complexity of  MAC protocols.  

In 2012, the IEEE 802.11ad standard \cite{11ad} was introduced as an amendment to IEEE 802.11 that enables bidirectional transmissions over the unlicensed 60 GHz mmW frequency band and support a variety of services with different QoS requirements. In addition, this standard supports fast session transfer (FST) that enables STAs to dynamically migrate from one frequency band to another. This capability will enable advanced multi-band STAs to jointly manage their traffic over either 2.4, 5, or 60 GHz unlicensed frequency bands. The performance of IEEE 802.11ad is studied in \cite{7870651,6451302,5956194}. The authors in \cite{7870651} analyze the performance of IEEE 802.11ad MAC protocol using a three-dimensional Markov chain model. In \cite{6451302}, a directional cooperative scheme is proposed for 60 GHz mmW communications which is shown to improve the system performance, compared with the standard IEEE 802.11ad. In \cite{5956194}, a throughput analysis of IEEE 802.11ad under different modulation schemes is presented. \textcolor{black}{The works in \cite{7870651,6451302,5956194} focus solely on performance analysis of the IEEE 802.11ad as a stand-alone system, although this standard has been designed to coexist with legacy IEEE 802.11.}



The main contribution of this paper is an integrated mmW-$\mu$W MAC protocol that enables STAs to dynamically leverage the bandwidth available at the $60$ GHz mmW band and alleviate the excessive delay caused by the contention-based medium access over the $\mu$W frequencies. In addition, we present a comprehensive performance analysis for the proposed protocol by adopting a Markov chain model for backoff time that accommodates FST between mmW and $\mu$W frequency bands. Furthermore, simulation results are provided and shown to perfectly corroborate the derived analytical results. Both analytical and simulation results show that the proposed MAC protocol significantly increases the saturation throughput and reduces the delay, compared with the legacy IEEE 802.11 DCF. Moreover, the impact of different network parameters, such as mmW link state, initial backoff window size, and maximum backoff stage on the performance are studied.

The rest of this paper is organized as follows. Section II presents the proposed MAC protocol. Section III presents the analytical results. Simulation results are provided in Section IV. Section V concludes the paper.   

\section{MAC Protocol Integration for Multi-Band Sub-6 GHz and MMW  WLANs}

The contention-based medium access in the IEEE 802.11 DCF suffers from increased backoff time and excessive delays in congested scenarios \cite{1204546,1357025}. To alleviate this problem, our goal is to leverage the multi-band operability of modern STAs to avoid excessive backoff times for collided frames and thus, decrease the associated contention delay for services in WLANs.
  \emph{ Prior to presenting the proposed scheme, we briefly overview some of the key definitions in the IEEE 802.11ad MAC protocol and 802.11 DCF that will be used in our analysis within the subsequent sections.}




\subsection{IEEE 802.11AD MAC Protocol Overview}\label{sec:IIA}
\begin{figure}[t!]
	\centering
	\centerline{\includegraphics[width=8cm]{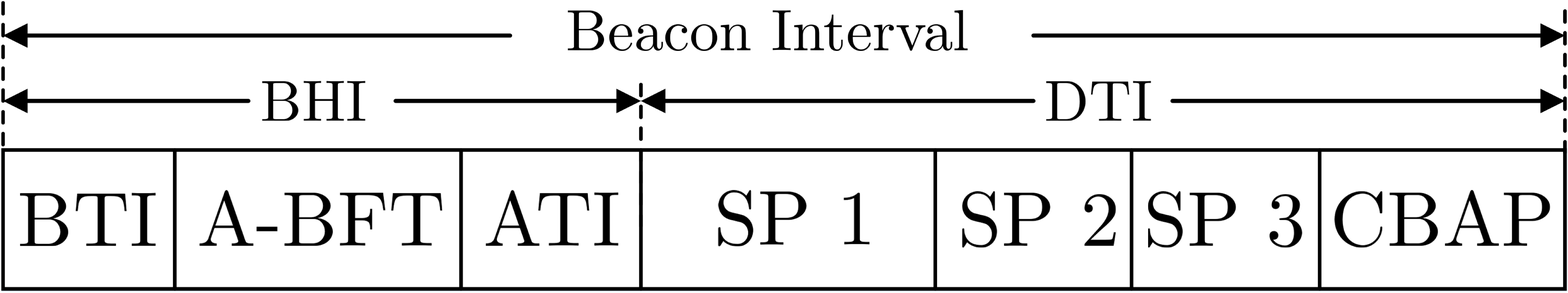}}\vspace{-0cm}
	\caption{\small Beacon Interval structure \cite{11ad}.}\vspace{-.2cm}
	\label{BI}
\end{figure} 
IEEE 802.11 standards, including IEEE 802.11ad, organize the medium access using periodic recurring beacon intervals (BIs). 
To accommodate bidirectional transmissions over the $60$ GHz mmW band, some adjustments are introduced in the IEEE 802.11ad BI structure, as shown in Fig. \ref{BI}. These modifications include:  1) sending directional beacon frames via an antenna sweeping mechanism, implemented within the beacon time interval (BTI). This sweeping process allows to extend the communication range and resolve the issue of STA discovery with unknown directions, 2) association beamforming training (A-BFT) used by stations to train their antenna sector for communication with the
personal basic service set (PBSS) control point (PCP)/access point (AP), and 3) the PCP/AP exchanges
management information, including scheduling, with beam-trained STAs prior to the data transmission interval (DTI). 

During DTI, three different medium access schemes are supported, namely, 1) contention-based access, 2) scheduled channel time allocation, and 3) dynamic channel time allocation. The first scheme which is conventional in IEEE 802.11 protocols allows STAs to access channel during  contention-based access periods (CBAPs). 
Two latter approaches are based on time devision multiple access (TDMA) that dedicate a service period (SP) to each pair of scheduled STAs. The dynamic channel time allocation method includes a polling phase (PP) that enables STAs to request a channel time from the PCP/AP. The PCP/AP allocates the available channel time according to these requests. This polling-based scheduling mechanism is implemented within the beacon header interval (BHI).

\vspace{-.1cm}
\subsection{IEEE 802.11 DCF Overview}\label{sec:IIB}
In this widely adopted protocol, STAs follow the contention-based carrier-sense multiple access with collision avoidance (CSMA/CA) scheme to reduce collisions \cite{ieee80211}. That is, an STA senses the channel prior to sending its packet. If channel is sensed busy, the STA defers the transmission until the channel is sensed idle for a DCF Interframe Space (DIFS) time. Afterwards, the STA chooses a random backoff counter (BC). Then, time is divided into slots and the BC will be decremented after each idle slot time. Moreover, the BC countdown is stopped, whenever the channel is sensed busy during a slot time. The BC count down is reactivated once the channel is sensed idle again for a DIFS. The STA sends its packet immediately after BC reaches zero.

The BC is randomly selected from integers within an interval [0, CW-1], where CW is called contention window. CW depends
on the number of transmissions failed for the packet. Initially, CW is set equal to a value $W$, called minimum contention window. After each unsuccessful transmission, $W$ is doubled, up to a maximum value of $2^mW$. At this point, if transmission fails again, the packet is either dropped or a new BC is chosen randomly from $[0,2^mW-1]$. 


\subsection{Proposed Integrated MmW-Microwave MAC Protocol}

\begin{figure}[t!]
	\centering
	\centerline{\includegraphics[width=\columnwidth]{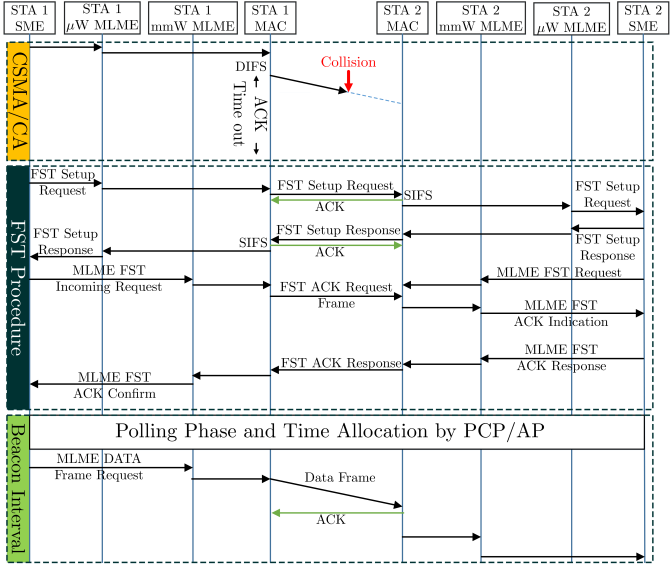}}\vspace{-0cm}
	\caption{\small Proposed Multi-Band MAC Protocol.}\vspace{-.3cm}
	\label{protocol}
\end{figure}

\textcolor{black}{In this work, we focus on the IEEE 802.11 DCF and IEEE 802.11ad dynamic channel time allocation, respectively, at the $\mu$W and mmW unlicensed bands.} In order to reduce the excessive delay caused by the collisions at the IEEE 802.11 DCF, in this section, we propose a novel protocol that enables STAs with multi-band capability to transfer their traffic to the contention-free $60$ GHz mmW band, whenever available, and avoid intolerable large backoff times. The proposed protocol is shown in Fig. \ref{protocol}. In this example scenario, STAs 1 and 2 are, respectively, the transmitting and receiving stations. The communications between STAs 1 and 2 can be explained in three following phases:

\textbf{Phase 1:} STA 1 aims to transmit its packet to STA 2, over the $\mu$W band using a CSMA/CA scheme, as explained in Sec. \ref{sec:IIB}. Due to its omnidirectional MAC protocol, the DCF of IEEE 802.11 requires minimum coordination among STAs, which provides a fast and flexible medium access. However, as the number of STAs increases, larger backoff times are required, resulting in more delay for packet transmissions. According to this protocol, STA 1 increases its backoff stage after each unsuccessful transmission.  After reaching the maximum backoff stage $m$, STA 1 initiates Phase 2 with probability $\beta$ and remains in Phase 1 with probability $1-\beta$. The merit of using this control parameter will be elaborated in the next section.

\textbf{Phase 2:} In this phase, STA 1 initiates an FST with STA 2. FST capability is introduced in the IEEE 802.11ad Extended version \cite{11ad} that enables STAs to swiftly move their traffic from one transmission band/channel to another. Since the FST is managed at a separate control channel, it will not be prone to collisions at the data channel. As shown in Fig. \ref{protocol}, to invoke FST, the station management entity (SME) unit in STA 1 sends an \emph{FST Setup Request} to the $\mu$W MAC layer management entity (MLME), followed by informing the STA 1's MAC to forward the \emph{FST Setup Request} frame to STA 2. Then, a handshaking procedure is done between STAs 1 and 2 in which STA 2 confirms that it is ready to move the communication to the $60$ GHz band. Up to this stage, the control messages between STAs 1 and 2 are exchanged at the $\mu$W band. Next, an \emph{FST ACK Request} is initiated by the STA 1's mmW MLME to request an FST ACK frame from STA 2. This message is transferred over the $60$ GHz band and FST is done once STA 1 receives the \emph{FST ACK Response} frame from STA 2. 

Here, we note that the FST procedure is revoked if STA 1 does not receive ACK frames in any stage during Phase 2. This can happen if the link between STAs 1 and 2 is blocked by an obstacle, or A-BFT is failed. In that case, STA 1 continues following the CSMA/CA in Phase 1.

\textbf{Phase 3:} This phase starts with the next BI of the IEEE 802.11ad, in which STA 1 participates in the polling within PP of BI and requests a contention-free time for communication with STA 2, as elaborated in Sec. \ref{sec:IIA}. Next, STA 1 will transmit its packet to STA 2 during the allocated SP in DTI. \textcolor{black}{Afterwards, STA 1 will reset its CW to the minimum value $W$ and it will initiate Phase 1.} 

The proposed multi-band MAC benefits from the flexible and simple CSMA/CA protocol at the $\mu$W unlicensed bands, while preventing excessive delays caused by the contention-based medium access. Next, we present analytical results to evaluate the performance of the proposed MAC protocol.

\section{Modeling and Analysis of the Proposed Multi-Band MAC Protocol}
In this section, we present analytical results to evaluate the performance of the proposed multi-band MAC protocol. First, we study the operation of an arbitrary STA that follows the proposed MAC protocol. In particular, we determine the probability of packet transmissions over either mmW or $\mu$W frequencies at a randomly chosen time slot. Then, we use these transmission probabilities to find a suitable expression for the saturation throughput.  
\subsection{Probability of Packet Transmission over mmW and $\mu$W Frequencies}
In our analysis, we assume non-empty queues for all STAs, i.e., the network operates at a saturation condition. As such, a new packet will be ready for transmission immediately after each successful transmission. These consecutive transmissions will require each STA \textcolor{black}{transmitting over the $\mu$W frequency band} to wait for a random backoff time prior to sending the next packet. In this regard, let $b(t)$ be the stochastic process for the BC of an arbitrary STA. A discrete and integer time scale is adopted  in which $t$  and $t+1$ present the beginning of two consecutive slot times, and the BC of each
STA is decremented at the beginning of each slot time. According to the works in \cite{840210} and \cite{4100720}, the DCF of IEEE 802.11 can be modeled as a two-dimensional discrete-time Markov chain $(s(t), b(t))$, where $s(t) \in \left\{0, 1, \cdots m\right\}$ represents the backoff stage of an STA at time $t$, with $m$ being the maximum backoff stage. \textcolor{black}{For an arbitrary backoff stage $s(t)=i$, the CW will be $W_i=2^iW$. In such Markov chain models, it is collectively assumed that, regardless of state, a collision occurs with a constant and independent probability $p$ as concretely discussed in \cite{840210} and~\cite{4100720}.} 
\begin{figure}[t!]
	\centering
	\centerline{\includegraphics[width=7cm]{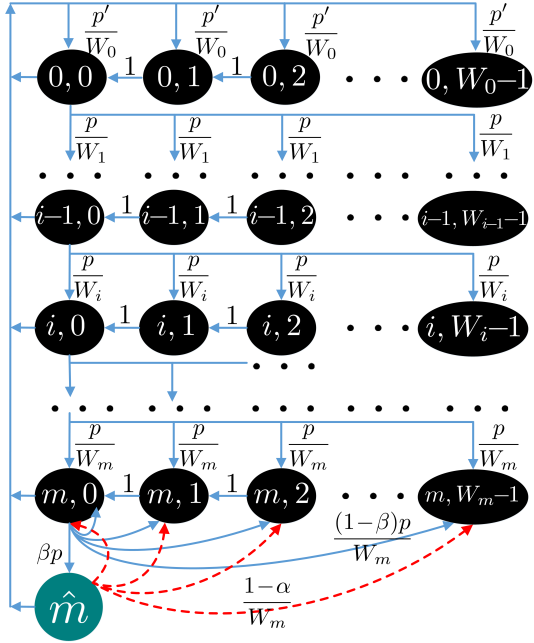}}\vspace{-0cm}
	\caption{\small Markov Chain model for the backoff window size}\vspace{-.2cm}
	\label{markov}
\end{figure}

To study the performance of the proposed protocol, we adopt a Markov chain model, as shown in Fig. \ref{markov}, where each state $(i,k)$ indicates that $s(t)=i$ and $b(t)=k$, i.e., the BC of an STA is at the $k$-th step of stage $i$. In addition, by introducing a new state $\hat{m}$, this model captures the capability of multi-band STAs to operate at the mmW frequency band. 
In fact, while being at state $(m,0)$, the STA can choose to either stay at the $\mu$W band and follow the DCF protocol or perform an FST to transmit over the mmW band. \textcolor{black}{We note that performing FST by an arbitrary STA $j$ does not alter the collision probability $p$ for other packets, since the next backlogged packet of STA $j$ will be ready to be sent over the $\mu$W frequency band}. In this model, $\beta \in [0,1]$ is a control parameter that allows an STA to manage unnecessary FSTs to reduce signaling overhead or avoid the mmW frequency band whenever \textcolor{black}{the transmission of a number of previous packets has failed, due to unsuccessful A-BFTs.}
 Moreover, $\beta$ provides backward compatibility for legacy STAs with no mmW communications capability\footnote{By choosing $\beta=0$, the proposed model will converge to the corresponding Markov chain for the conventional DCF in IEEE 802.11 standards.}. The state of each mmW link is determined by a Bernoulli random variable $\eta$ with success probability $\alpha$. That is, \textcolor{black}{with probability $\alpha_j$, a transmitting STA $j$ and its desired receiving STA can successfully perform the A-BFT and execute the transmission.} Here, the single-step nonzero transition probabilities are
\begin{subequations}
	\begin{IEEEeqnarray}{rCl}
&&\!\!\!\!\!\!\!\!\!\mathbbm{P}\left\{i,k | i,k+1 \right\}=1,\,\,\,\,\,\,\,\,\,\,\,\, \,\,\,\,\,\, \hfill i \in [0,m], k\in [0,W_i\!-\!2],\label{1a}\\
&&\!\!\!\!\!\!\!\!\!\mathbbm{P}\left\{0,k | i,0 \right\}=p'/W_0,  \hfill i \in [0,m]\cup \left\{\hat{m}\right\}, k\in [0,W_0\!-\!1], \label{1b}\\
&&\!\!\!\!\!\!\!\!\!\mathbbm{P}\left\{m,k | \hat{m} \right\}=(1-\alpha)/W_m, \hfill  k\in [0,W_m\!-\!1],\label{1c}\\
&&\!\!\!\!\!\!\!\!\!\mathbbm{P}\left\{i,k | i-1,0 \right\}=p/W_i, \hfill   i \in [1,m], k\in [0,W_i\!-\!1],\label{1d}\\
&&\!\!\!\!\!\!\!\!\!\mathbbm{P}\left\{m,k | m,0 \right\}=p(1-\beta)/W_m, \hfill  k\in [0,W_m\!-\!1],\label{1e}\\
&&\!\!\!\!\!\!\!\!\!\mathbbm{P}\left\{\hat{m}| m,0 \right\}=\beta p, \label{1f}
\end{IEEEeqnarray}
\end{subequations}
where \eqref{1a} shows the backoff time count down at each time slot. Moreover, \eqref{1b} indicates that, after a successful packet transmission, an STA will randomly choose a BC from stage $0$, i.e., $k$ is uniformly chosen from $[0,W-1]$. In \eqref{1b}, $p'$ is equal to $1-p$ and $\alpha$, respectively, if $i\in [0,m]$ and $i=\hat{m}$. In addition,  \eqref{1c} captures an unsuccessful mmW transmission, after which the STA will remain at the $\mu$W frequency band and will choose a random backoff time at stage $m$. \eqref{1d} shows that backoff stage will incremented after an unsuccessful $\mu$W transmission. Furthermore, \eqref{1e} and \eqref{1f} indicate, respectively, that an STA will remain at stage $m$ with probability $1-\beta$ after a collision, or will perform an FST with probability $\beta$.

For this Markov chain model, we next determine the stationary probability for each state $(i,k)$. Let $h_{i,k}=\lim\limits_{t \rightarrow \infty}\mathbbm{P}\{s(t)=i,b(t)=k\}$, $i\in[0,m]\cup\{\hat{m}\}$, $k\in[0,W_i-1]$. From the Markov chain model in Fig. \ref{markov}, it is easy to see that
\begin{align}\label{eq2}
	h_{i,0} =ph_{i-1,0}=p^i h_{0,0},\hspace{0.5cm} i\in(0,m).
\end{align}
Furthermore, for $i=m$ and $\hat{m}$, we note that
\begin{subequations}
	\begin{IEEEeqnarray}{rCl}
&&ph_{m-1,0} + p(1-\beta)h_{m,0} + (1-\alpha)h_{\hat{m}} = h_{m,0},\label{3a}\\
&&p\beta h_{m,0} = h_{\hat{m}}.\label{3b}
\end{IEEEeqnarray}
\end{subequations}
Using \eqref{eq2}, we solve \eqref{3a} and \eqref{3b} with respect to $h_{m,0}$ and $h_{\hat{m}}$, which yields 
\begin{align}\label{eq:4}
	h_{m,0} = \frac{p^m}{1-p+\alpha \beta p} h_{0,0}, \hspace{.2cm} h_{\hat{m}}=\frac{\beta p^{m+1}}{1-p+\alpha \beta p}h_{0,0}.
\end{align}
Next, by following the chain regularities, we can represent the remaining stationary state probabilities as:
\begin{align}\label{utility1}
h_{i,k}=W'_i
&\begin{cases}
\frac{1-p}{W_0}\sum_{j=0}^{m}h_{j,0}+\frac{\alpha}{W_0}h_{\hat{m}},\hspace{1.85cm} i=0,\\
\frac{p}{W_m}h_{m-1,0} + \frac{p(1-\beta)}{W_m}h_{m,0} + \frac{1-\alpha}{W_m}h_{\hat{m}},i=m,\\
\frac{p}{W_i}h_{i-1,0}, \hspace{3.35cm} i\in (0,m),
\end{cases}
\end{align}
where $W'_i=W_i-k$, and $k \in (0,W_i-1]$. In addition, we note that
\begin{align}\label{eq:h_0,0}
	h_{0,0}= (1-p)\sum_{j=0}^{m}h_{j,0}+\alpha h_{\hat{m}}.
\end{align}
Thus, by using \eqref{eq2}, \eqref{3a}, and \eqref{eq:h_0,0}, $h_{i,k}$ in \eqref{utility1} simplifies to:
\begin{align}\label{eq:h_i,k}
	h_{i,k}=\frac{W_i-k}{W_i}h_{i,0},\,\,\,\,\,\,\,\,\,\,\, \hfill i\in [0,m], k\in (0,W_i-1].
\end{align}
Finally, we find $h_{0,0}$ by noting that the sum of all state probabilities is $1$. That is,
\begin{align}\label{eq:h00-1}
	1&=\sum_{i=0}^{m}\sum_{k=0}^{W_i-1}h_{i,k} + h_{\hat{m}}\\
	&\stackrel{\text{(a)}}{=}\sum_{i=0}^{m}h_{i,0}\sum_{k=0}^{W_i-1}\frac{W_i-k}{W_i}+h_{\hat{m}},\notag\\
	&\stackrel{\text{(b)}}{=}\sum_{i=0}^{m-1}h_{i,0}\frac{W_i+1}{2}+\frac{W_m+1}{2}b_{m,0}+h_{\hat{m}},\notag\\
	&\stackrel{\text{(c)}}{=}\left[\sum_{i=0}^{m-1}(W_i+1)p^i \!+\!\frac{(W_m+1)p^m}{1-p+\alpha \beta p}\!+\!\frac{2\beta p^{m+1}}{1-p+\alpha \beta p}\right]\frac{h_{0,0}}{2} .\notag
\end{align}
In \eqref{eq:h00-1}, (a) and (b) result from \eqref{eq:h_i,k} and noting that $\sum_{k=0}^{W_i-1}(W_i-k)/{W_i}=(W_i+1)/2$, respectively. In addition, (c) results from \eqref{eq2} and \eqref{eq:4}. From \eqref{eq:h00-1}, we can find $h_{0,0}$ as follows:
\begin{align}\label{h00}
	h_{0,0}&=2\left[W\left(\frac{1-(2p)^m}{1-2p}\right)\right.\notag\\
	& +\left.\frac{1-p^m}{1-p}+\frac{(2^mW+1+2\beta p)p^m}{1-p+\alpha \beta p}\right]^{-1}.
\end{align}
Next, we can compute the transmission probability over the $\mu$W band, $\Theta^{\mu\text{W}}$, for an STA in a random time slot. To this end, we note that $\mu$W transmission occurs only if the backoff time countdown for an STA reaches zero. That is, an STA transmits a packet if it is at any states $(i,0), i\in [0,m]$. Thus,
\begin{align}\label{eq:theta}
	\Theta^{\mu\text{W}} = \sum_{i=0}^{m}h_{i,0}
	=\frac{1}{1-p}\left[1-\frac{\alpha \beta p^{m+1}}{1-p+\alpha \beta p}\right]h_{0,0}.
\end{align}
\begin{remark}
	Without mmW communications ($\beta=0$), we can easily verify that $\Theta^{\mu\text{W}}$ in \eqref{eq:theta} simplifies to
	\begin{align}
			\Theta^{\mu\text{W}} =\frac{2(1-2p)}{(1-2p)(W+1)+pW(1-(2p)^m)},
	\end{align}
	which is shown to be the transmission probability in DCF protocol of the IEEE 802.11  \cite{840210}.
\end{remark}
Over the mmW frequency band, STAs that are in state $\hat{m}$ will be scheduled to transmit within the next available DTI. Given that the mmW transmissions follow a TDMA scheme during each SP, as proposed in IEEE 802.11ad, no collision will happen. However, as mentioned in section \ref{sec:IIA}, a mmW transmission is contingent upon a successful A-BFT phase. Hence, the probability of transmission over the mmW frequency band is 
\begin{align}
\Theta^{\text{mmW}} =\mathbbm{P}\{\eta=1\} h_{\hat{m}}=\frac{\alpha\beta p^{m+1}}{1-p+\alpha \beta p}h_{0,0}.
\end{align}
After deriving the transmission probability at both mmW and $\mu$W frequencies for an arbitrary STA, our next step is to compute the saturation throughput as a key performance~metric.
\vspace{-.8cm}
\subsection{Throughput Analysis of the Proposed Multi-band mmW-$\mu$W MAC Protocol }
Next, we analyze the  system throughput $R$ at the saturation conditions. This throughput is defined as the average payload that is successfully transmitted across the network during a randomly chosen time slot, divided by the average time slot duration $\mathbbm{E}[T]$. 
In multi-band WLANs, parallel streams of data can be sent simultaneously over different frequency bands. Thus, our analysis will focus on finding  the throughput across the aggregated mmW-$\mu$W frequencies. 

Consider a WLAN, composed of $J$ STAs within a set $\mathcal{J}$. Over the $\mu$W frequency band, the protocol follows the standard CSMA/CA. In other words, only one STA can successfully transmit at a given time, otherwise, collision happens. In this regard, $P^{\mu\text{W}}_t$ is defined as the probability that at least one STA is transmitting over the $\mu$W frequency band. Since each STA $j \in \mathcal{J}$ transmits with probability $\Theta_j^{\mu\text{W}}$, $P^{\mu\text{W}}_t$ is given by:
 \begin{align}\label{eqPT1}
 	P_t^{\mu\text{W}} = 1-\prod_{j\in \mathcal{J}}(1-\Theta_j^{\mu\text{W}}).
 \end{align}  
In addition, transmission of an arbitrary STA $j$ is successful, if no other STA transmits at the same time. Hence, the probability of successful transmission can be written as: 
 \begin{align}\label{eqPT2}
P_s^{\mu\text{W}} = \frac{\sum_{j \in \mathcal{J}}\Theta_{j}^{\mu\text{W}}\prod_{j'\in \mathcal{J}\backslash j}(1-\Theta_{j'}^{\mu\text{W}})}{P_t^{\mu\text{W}}}.
\end{align} 
To compute $\mathbbm{E}[T]$, we note that there are three possible cases for the transmission scenarios over the $\mu$W band: 1) having an empty slot which occurs with probability  $1-P^{\mu\text{W}}_t$, since no STA is transmitting. 2) Successful transmission of a packet during a time slot which happens with probability $P^{\mu\text{W}}_t P^{\mu\text{W}}_s$, and 3) collision scenario that occurs with probability  
$P^{\mu\text{W}}_t(1-P^{\mu\text{W}}_s)$. Hence, the average slot time is  
\begin{align}
\!\!\!\mathbbm{E}[T]\!=\!(1\!-\!P^{\mu\text{W}}_t)\sigma \!+\! P^{\mu\text{W}}_t P^{\mu\text{W}}_s T_s\!+\! P^{\mu\text{W}}_t(1-P^{\mu\text{W}}_s)T_c,
\end{align}
where $T_s$, $T_c$, and $\sigma$ denote the slot time duration, respectively, in successful, collision, and no transmission scenarios.

For mmW transmissions, we must note that only FST is performed over the $\mu$W band, while other phases during BHI as well as payload transmissions in DTI will be done simultaneously with the $\mu$W band transmissions. To properly capture the mmW band contribution in the system throughput, we consider the time overhead associated with performing FST and we find the average number of STAs that can be scheduled at the mmW frequency band within a coarse of $\mathbbm{E}[T]$ time.  In this regard, let $\hat{J}\leq J$ be the maximum number of STAs that can be scheduled over the mmW band during $\mathbbm{E}[T]$, each transmitting a payload of size $B^{\text{mmW}}$ bits. Considering $r^\text{mmW}$ as the mmW channel bit rate, $\hat{J}=\lfloor \mathbbm{E}[T]r^\text{mmW}/B^{\text{mmW}}\rfloor$, where $\lfloor . \rfloor$ is the floor operand. Consequently, the average number of STAs transmitting at the mmW frequency band, $	\mathbbm{E}[J^{\text{mmW}}]$,  is
\begin{align}\label{eqPT3}
	\mathbbm{E}[J^{\text{mmW}}]=\sum_{u=1}^{\hat{J}}\sum_{s=1}^{J \choose u}\prod_{j=1}^{|\mathcal{J'}|=u} \Theta_j^{\text{mmW}},
\end{align}
where the inner sum is done over all possible subsets $\mathcal{J'} \subseteq\mathcal{J}$ with $|\mathcal{J'}|=u$ number of STAs. Clearly, there are $J \choose u$ distinct subsets with size $u$. Moreover, the product is for all STAs in the chosen subset $\mathcal{J'}$. In addition, since the protocol employs TDMA scheme for mmW communications, no collision will occur between multiple mmW transmissions during a DTI and the probability of successful transmission is $P_s^{\text{mmW}}=1$.
 
	\begin{table}[!t]
	\scriptsize
	\centering
	\caption{
		\vspace*{-0cm}Simulation parameters}\vspace*{-0.2cm}
	\begin{tabular}{|c|c|c|}
		\hline
		\bf{Notation} & \bf{Parameter} & \bf{Value} \\
		\hline
		$H_{\text{MAC}}$ & MAC header & $272$ bits\\
		\hline
		$H_{\text{PHY}}$ & PHY header & $128$ bits\\
		\hline
		$B^{\mu\text{W}}$ & $\mu$W packet payload & $8184$ bits\\
		\hline
		$B^{\text{mmW}}$ & mmW packet payload & $81840$ bits\\
		\hline
		ACK & ACK & $112$ bits + PHY header\\
		\hline
		$\delta$ & Propagation delay & $1$ $\mu$s\\
		\hline
		$\sigma$ & Slot time & $50$ $\mu$s\\
		\hline
		SIFS & Short interframe
		space & $28$ $\mu$s\\
		\hline
		DIFS & Distributed
		interframe space & $128$ $\mu$s\\
		\hline
		  $r^{\mu\text{W}}$ & $\mu$W channel bit rate&$1$ Mbps\\
		\hline
		  $r^\text{mmW}$ & mmW channel bit rate&$1$ Gbps\\
		\hline
		\text{SETUP\textunderscore REQ}& FST setup request& 240 bits\\
		\hline
		\text{SETUP\textunderscore RES}& FST setup response&240 bits\\
		\hline
	\end{tabular}\label{tabsim}\vspace{-1em}
\end{table}

Therefore, the system throughput $R$ is calculated by finding the aggregated transmitted payload over both $\mu$W and mmW frequency bands, divided by the average time slot duration $\mathbbm{E}[T]$ plus the time overhead associated with FST process:
\begin{align}
R = \frac{P_s P_t B^{\mu\text{W}}+\mathbbm{E}[J^{\text{mmW}}]B^{\text{mmW}}}{\mathbbm{E}[T]+\mathbbm{E}[J^{\text{mmW}}] T_{\text{FST}}},
\end{align}
where $B^{\mu\text{W}}$ is the payload size over the $\mu$W frequency band. Given the high available bandwidth at the mmW band, $B^{\text{mmW}}$ is considered larger than $B^{\mu\text{W}}$. Moreover, $T_{\text{FST}}$ is the required time for performing an FST. 


\vspace{-.5em}

\section{Simulation Results}
We validate our analytical results by simulating the proposed protocol in a multi-band WLAN. The number of STAs varies from $J=5$ to $50$. The considered network is simulated in MATLAB and the total simulation time extends to $500$ seconds. We consider $\alpha_j\!=\!\alpha, j\in \mathcal{J}$ to simplify the performance analysis. In this case, \eqref{eqPT1}-\eqref{eqPT3} can be written as:
 \begin{subequations}
 	\begin{IEEEeqnarray}{rCl}\label{eqPT4}
&&P_t^{\mu\text{W}} = 1-(1-\Theta^{\mu\text{W}})^J,\\
&&P_s^{\mu\text{W}} = {J\Theta^{\mu\text{W}}(1-\Theta^{\mu\text{W}})^{J-1}}/{P_t^{\mu\text{W}}},\\
&&\mathbbm{E}[J^{\text{mmW}}]=\sum_{u=1}^{\hat{J}}   {{J}\choose{u}}\left(\Theta^{\text{mmW}}\right)^u.
\end{IEEEeqnarray}
\end{subequations}
The effect of $\alpha$ and $\beta$ on the network performance will be evaluated subsequently. For $\mu$W communications, we consider the basic access scheme\footnote{Other access schemes such as request-to-send/clear-to-send (RTS/CTS) mechanisms can be applied similarly.} in which the receiving STA will send an acknowledgment (ACK) signal after successfully decoding the sent packet. Hence, $T_s$, $T_c$ and $T_\text{FST}$ are calculated as~follows:
\begin{align}
	&T_s = \Gamma+\text{SIFS}+\text{ACK}+\text{DIFS}+2\delta,\notag\\
	&T_c = \Gamma+\text{DIFS}+\delta,\notag\\
	&T_{\text{FST}} =\text{SETUP\textunderscore REQ} + \text{SETUP\textunderscore RES}+2\text{ACK}+4\delta,
\end{align}
where $\Gamma$ is the required time for transmitting PHY header $H_{\text{PHY}}$, MAC header $H_{\text{MAC}}$, and payload $B^{\mu\text{W}}$ of a $\mu$W packet. Moreover, $\delta$ models the propagation delay. $T_\text{FST}$ is calculated based on the FST procedure, as shown in Fig. \ref{protocol}, composed of sending FST Setup Request/Response frames, each followed by an ACK signal. Here, we note that FST ACK Request/Response frames are sent over the mmW frequency band, thus, they are not included in the time overhead. All network parameters are summarized in Table~\ref{tabsim}.

\begin{figure}[t!]
	\centering
	\centerline{\includegraphics[width=8cm]{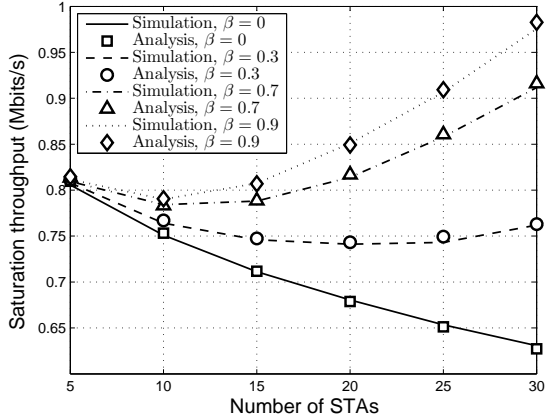}}\vspace{-.2cm}
	\caption{\small Saturation throughput vs the number of STAs.}\vspace{-.3cm}
	\label{fig1}
\end{figure} 

Fig. \ref{fig1} shows the effect of the control parameter $\beta$ on the performance, for different number of STAs, with $m=3$, $W=32$, and $\alpha=0.6$. From Fig. \ref{fig1}, we can see that the throughput increases as $\beta$ becomes large. Interestingly, this performance gain is more evident for large network sizes, since the collision probability is higher and, thus, the proposed protocol sends more packets over the mmW band. In addition, for any fixed non-zero $\beta$, we observe that the throughput initially decreases and then increases, as the number of STAs grows. That is because, for a larger network size $J$, collisions initially increase which results in a lower throughput. However, for larger network sizes, more STAs reach the maximum backoff stage $m$ and initiate FST to the mmW band.

\begin{figure}[t!]
	\centering
	\centerline{\includegraphics[width=8cm]{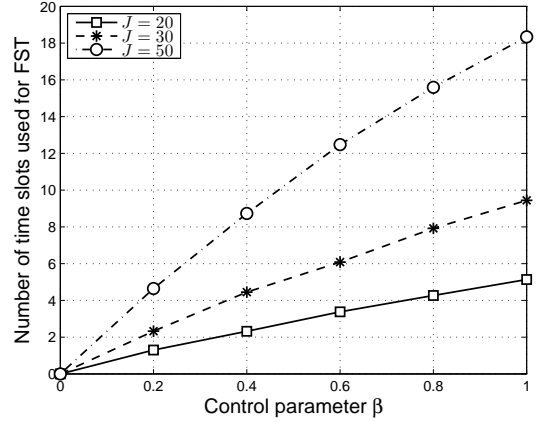}}\vspace{-.2cm}
	\caption{\small Number of time slots used in FST procedure vs the control parameter $\beta$, for different network size $J$.}\vspace{-.55cm}
	\label{fig9}
\end{figure} 

In Fig. \ref{fig9}, the overhead of the proposed protocol is evaluated in terms of the number of time slots used in the FST procedure. From Figs. \ref{fig1} and \ref{fig9}, we observe an interesting tradeoff between the saturation throughput and the overhead of switching between mmW and $\mu$W frequency bands. For example, from Fig. \ref{fig1}, we can see that the throughput is improved by $28\%$ for $J=30$, when $\beta$ is increased from $\beta=0.3$ to $\beta=0.9$. Moreover, Fig. \ref{fig9} shows that the overhead increases from $3$ slots to $9$ slots in order to achieve this performance gain. 

\begin{figure}[t!]
	\centering
	\centerline{\includegraphics[width=8cm]{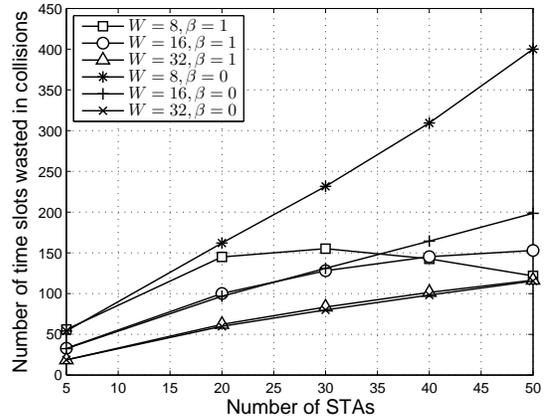}}\vspace{-.2cm}
	\caption{\small Number of time slots wasted in collisions vs the number of STAs, for different $W$ and $\beta$ values.}\vspace{-.3cm}
	\label{fig6}
\end{figure}

Fig. \ref{fig6} shows another key merit of mmW-$\mu$W MAC layer integration which is reducing the packet transmissions delay caused by the collisions. This figure compares the number of time slots that are wasted in collisions by the proposed protocol ($\beta=1$) and legacy IEEE 802.11 ($\beta=0$), for different initial contention window and network sizes. From Fig. \ref{fig6}, it is clear that the proposed scheme significantly reduces the delay, e.g., up to three times for $J=50$ STAs and $W=8$.
Moreover, we observe that the performance gap between the two schemes is larger for smaller $W$ values. That is because more collisions occur when initial backoff window size is small, which increases the probability for STAs to transmit their packets over the mmW frequency band. 

\begin{figure}[t!]
	\centering
	\centerline{\includegraphics[width=8cm]{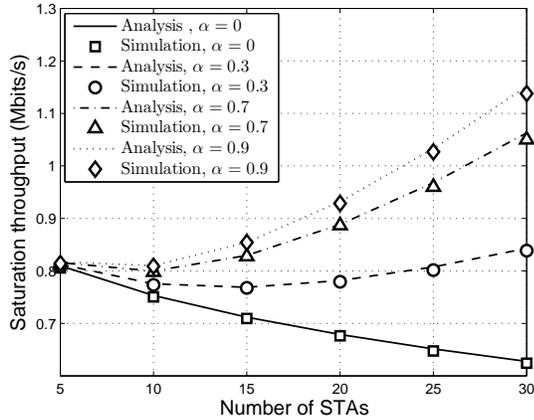}}\vspace{-.3cm}
	\caption{\small Saturation throughput vs the number of STAs.}\vspace{-.5cm}
	\label{fig2}
\end{figure} 

Fig. \ref{fig2} shows the saturation throughput as a function of $\alpha$ for different number of STAs, with $m=3$, $W=32$, and $\beta=1$. We can observe that, as mmW communication is more feasible, the throughput will increase with all network sizes. For example, the throughput increase by $37\%$ for $J=20$ and $\alpha=0.9$, compared with the stand-alone IEEE 802.11 system ($\alpha=0$). Similar to Fig. \ref{fig1}, the throughput varies as a convex function with respect to the number of STAs.

\begin{figure}[t!]
	\centering
	\centerline{\includegraphics[width=8cm]{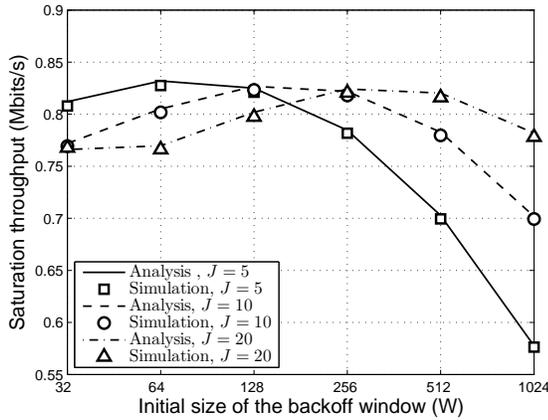}}\vspace{-.3cm}
	\caption{\small Saturation throughput vs $W$ for different network size $J$.}\vspace{-.3cm}
	\label{fig3}
\end{figure}

In Fig. \ref{fig3}, the impact of initial backoff window size, $W$, on the throughput is studied for $m=3$, $\alpha=\beta=0.5$, and three network sizes $J=5,10,20$. Fig. \ref{fig3} also shows the optimal $W$ for maximizing the throughput. We can observe that the optimal $W$ grows as the number of STAs $J$ increases. 
\begin{figure}[t!]
	\centering
	\centerline{\includegraphics[width=8cm]{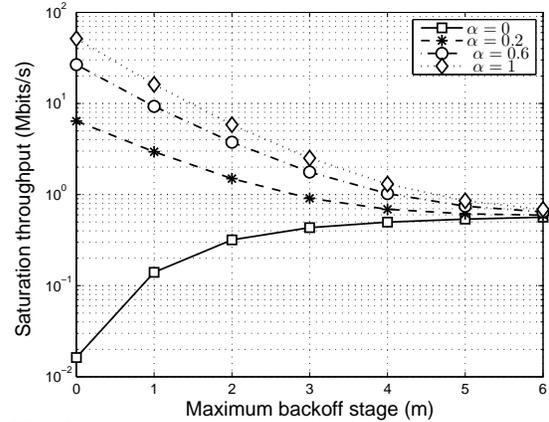}}\vspace{-0.3cm}
	\caption{\small Saturation throughput vs $m$ for different $\alpha$ values.}\vspace{-.3cm}
	\label{fig5}
\end{figure} 

Furthermore, the effect of maximum backoff stage, $m$, on throughput is shown in Fig. \ref{fig5} with $\beta=0.5$, $W=16$, and $J=50$. It is interesting to note that for $\alpha=0$, i.e., with no mmW communications, throughput increases as $m$ grows. That is because less collisions happen with larger maximum backoff. However, this trend is opposite for nonzero $\alpha$ values. In fact, even for $\alpha=0.2$ and small $m$, we observe a significant performance gain which results from STAs' frequent switching to the mmW frequency band, due to the high collision at the $\mu$W frequency band. 
\vspace{-0.25cm}
\section{Conclusions}\vspace{-.1cm}
In this paper, we have proposed a novel MAC protocol that leverages the capability of advanced wireless stations to decrease the contention-based delay and increase throughput in WLANs. In fact, the proposed protocol allows stations to perform fast session transfer to the 60 GHz mmW band, and avoid excessive delay caused by collisions at the $\mu$W unlicensed bands. To analyze the performance of the proposed scheme, we have adopted a Markov chain model that captures the fast session transfer across mmW-$\mu$W bands. We have shown the accuracy of the model by providing comprehensive simulation results. Both simulations and analytical results have shown that the proposed protocol yields significant gains in terms of maximizing the saturation throughput and minimizing the delay caused by collisions.

\vspace{-.2cm} 
\def\baselinestretch{.87}
\bibliographystyle{IEEEbib}
\bibliography{references}
\end{document}